# Integrating Newer Technology Software Systems into the SLAC Legacy Control System - Two Case Histories and New CMLOG Developments

M. Laznovsky and R. MacKenzie, Stanford Linear Accelerator Center, Stanford, CA 94309, USA
J. Chen, Thomas Jefferson National Accelerator Facility, Newport News, Virginia 23606, USA

Abstract

It has been the goal of SLAC Controls Software to offload processing from the aging Alpha/VMS based control system onto machines that are more widely accepted and used. An additional goal has been to provide more modern software tools to our user community. This paper presents two software products which satisfy those goals.

## 1 CMLOG: SLC INTEGRATION AND NEW DEVELOPMENTS

Common Message Logger (CMLOG)[1] is an already-existing object-oriented and distributed system that allows the logging of messages to a centralized database and lets users and applications view incoming messages in near-real-time using a Motif browser and retrieve stored data from the database. CMLOG was integrated with the legacy system by writing programs on VMS and Solaris which took advantage of the CMLOG Applications Programming Interface (API) for shipping messages (in near-real-time) from VMS to the CMLOG server running on Solaris. CMLOG is also used to log messages from EPICS IOCs to the CMLOG server. Those messages are forwarded from Solaris to the VMS system using the CMLOG API. New developments for CMLOG are presented. These in include a JAVA Browser.

### 1.1 SLC Integration

CMLOG provides an easy to use "Browser" API for retrieving messages from the cmlogServer. For the purpose of SLC integration, the forwarding program uses this API to retrieve messages in "Update" mode which provides the messages as they arrive at the cmlogServer from EPICS IOCs and other clients. Then, the forwarding program simply sends those messages on to the Alpha/VMS error logging system using a simple tcp/ip connection to an already-existing SLC interface.

It was necessary to write a TCP/IP server to receive messages originating on the Alpha/VMS Control System. This simple server receives messages from a sending program on Alpha/VMS and uses the CMLOG Client API to put the messages into the cmlogServer.

The result is that all the messages that exist in the legacy Alpha/VMS error log system also exist on the more modern CMLOG based system. This is seen as a temporary situation while the operators migrate to CMLOG.

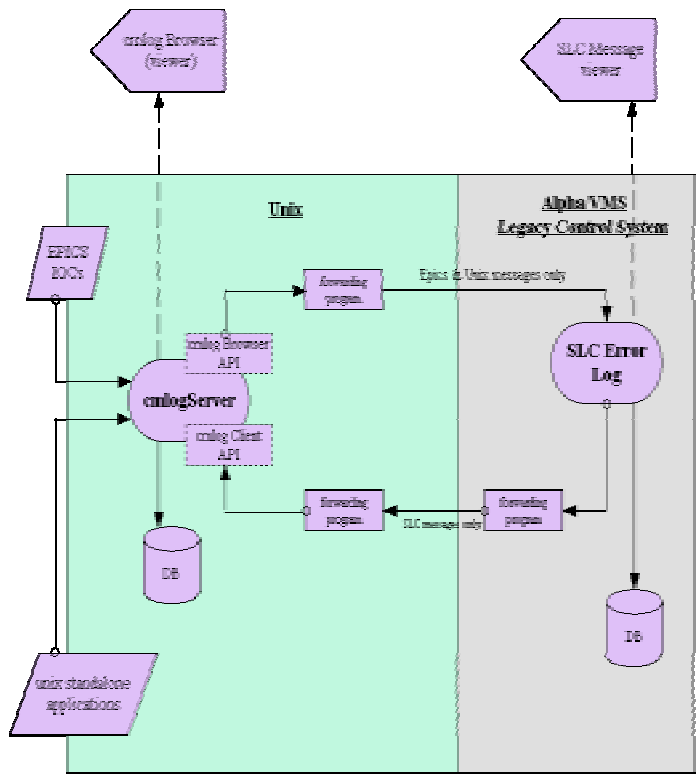

FIGURE 2.1: simple cmdSrv example configuration.

## 1.2 A New Generation of Message Throttling

The CMLOG client code was modified at SLAC to support a different style of message throttling than the one originally built into the CMLOG client. This new code is backwardly compitable with the old style of throttling (the old calls still work). Both C and C++ API calls are supported.

The original style CMLOG throttling maintains a history of unique field values and throttles on each of those separately. No provision is made to specify specific values of special interest for throttling; the system would just allow a certain number of each value through.

The new throttling calls support filtering on specific contents of tags (fields) of a message. A throttle "limit" specifies the number of messages desired per time period ("deltaTime"), and throttles may be re-started at any time with different limit and time-period parameters.

In addition, summary messages are output when throttles are started, stopped, and after throttling has resulted in messages being dropped. A function is provided for displaying the current list of throttles.

Here is an example of a call to set a new-style throttle. This call sets a throttle on the field(or tag) "text", allowing two messages every ten seconds which have the subsrtring "agv" in them.

```
filter -> setThrottle ("text",
                       2,
                       10.0,
                       "agv");
```

## 1.3 Java API and Browser.

The CMLOG Java API provides the same capability as the C++ and C Browser APIs. That is, programmers can write their own browser and other applications to fetch data from cmlogServer. The Java API is implemented as Java package 'cmlog'. The API is well documented on the web in standard javadoc format [3]. Useful example programs are provided as part of the standard CMLOG distribution.

The CMLOG Java Browser is implemented using the Java Swing package and makes use of the Java Browser API. It is supported on Java 1.2 and above. The java bytecode can be built from scratch using the CMLOG distribution, or, a pre-compiled file "cmlog.jar" is also provided.

The Java Browser has most of the same features that the Motif browser offers with the exception of: displaying error codes in different colors, displaying strings for a set of integers, and invoking user scripts. Those features will be added in future releases.

## 1.4 32 Bit Support

The latest release of CMLOG is 64 bit compatable. It has been tested on 64-bit solaris and Alpha Linux.

## 2 COMMAND SERVER

The Command Server (cmdSrv) [2] program currently runs on Solaris and accepts commands from remote nodes like the VMS legacy system and executes them. Functionally, cmdSrv is similar to UNIX rsh with additional capabilities including: increased performance, added security, ease of configuration, on-demand reconfiguration, static load balancing, a flexible client-side-API, and a query capability for keeping track of who executed what from where. cmdSrv is used for invoking many applications including EPICS dm displays and CMLOG. CmdSrv was designed primarily to execute X-displays but can also be used to execute other unix commands.

## 2.1 Features, Architecture and Design

The parent-process portion of cmdSrv is a typical tcp/ip server. It accepts new connections from remote client processes. For each connection, it forks a child process which, in turn, execs the command that was sent from the client. The parent process then waits a number of seconds to see if the child dies prematurely (probably due to an error in the command that was requested). If the child does die prematurely, then an error condition is returned to the client. Otherwise, success is returned to the client. Then, the connection is closed. This Architecture is shown in Figure 2.1

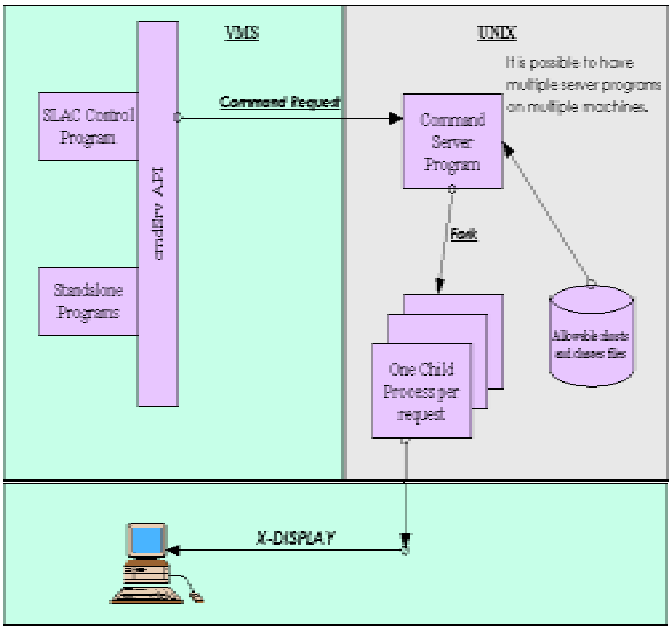

FIGURE 2.1: simple cmdSrv example configuration.

CmdSrv supports the following client command requests (sent inside the command request packet):
    EXEC - execute an x-command or script
    REFRESH - reload the control files.
    SHOW - return a list of active child processes.
    KILL - kill child processes.

The client command request packet also lets the client specify the command class for the command being executed. Only one command class is allowed per display head. This is useful for limiting the number of active x-displays that can be running in order to preserve resources.

There are two files which control the operation of cmdSrv. One file contains a list of network nodes from which client requests may come from. These are the only nodes from which commands may originate. The second file is a list of allowable command classes. Either file may be re-loaded into cmdSrv by sending the REFRESH command from any client client.

The cmdSrv parent process maintains a list of active child processes. For each child, the Requesting User, Process, Class, and Node are stored. Additionally, the X-display, PID, and requested command are stored. Any client may request to be sent this list of processes by sending the SHOW command to cmdSrv. This is very useful for remotely keeping track of what is running under cmdSrv and where any X-displays are active.

A flexible and portable "C" API is provided for client programmers. The api makes it easy to send command request messages from client programs (example programs are provided).

A flexible log file capability is available where the cmdSrv parent process logs (to a file) extensive information about each request that it receives from clients.